\documentclass[preprint,amsmath,amssymb,aps,prb]{revtex4-1}
\usepackage{graphicx}
\usepackage[colorlinks=true, allcolors=blue]{hyperref}
\usepackage{float}
\usepackage{booktabs}
\usepackage{lineno} 

\begin{document}
%\linenumbers

\title{Establishing simple relationship between eigenvector and matrix elements}
\author{W. Pan, J. Wang and D. Y. Sun}
\affiliation{Department of Physics, East China Normal University, 200241 Shanghai, China}

\begin{abstract}
A simple approximate relationship between the ground-state eigenvector and the sum of matrix elements in each row has been established for real symmetric matrices with non-positive off-diagonal elements. Specifically, the $i$-th components of the ground-state eigenvector could be calculated by $(-S_i)^p+c$, where $S_i$ is the sum of elements in the $i$-th row of the matrix with $p$ and $c$ being variational parameters. The simple relationship provides a straightforward method to directly calculate the ground-state eigenvector for a matrix. Our preliminary applications to the Hubbard model and the Ising model in a transverse field show encouraging results.The simple relationship also provide the optimal initial state for other more accurate methods, such as the Lanczos method.\\
\\

\end{abstract}
\maketitle

\section{Introduction}
Quantum many-body systems contain novel physical phenomena, which is the key research object in condensed matter physics. Physically, the diagonalization of Hamiltonian matrix is indispensable for the accurate treatment of quantum many-body problems. However, as the number of particles increases, scientists have to face the exponentially increased dimension of Hamiltonian matrices. Over years, physicists have made great efforts to handle this kind of matrices, and have reached fruitful achievements. To list just a few of them for instance, exact diagonalization method \cite{Lanczos1950,PhysRevLett.72.1545,Zhang2010,Si1994PRL,RevModPhys.68.13}, quantum Monte-Carlo \cite{Sandvik1991PRB,vonderLinden1992,PROKOFEV1998,Foulkes2001RMP,Kolorenc2011,RevModPhys.83.349,Gubernatis2016}, and the density matrix renormalization group \cite{White1992PRL,Schollwuck2005RMP,RevModPhys.80.395,Schollwuck2011AP,Verstraete2008}, $etc$. Each of them has been successfully applied to a vast of physical problems.

As far as we know, it is still lack of a general and efficient computational method to solve many-body problems, especially for strong correlated systems. Currently, it is safe to say that, the development of computational methods is one of most urgent task for the theoretical study of quantum many-body systems. Usually, two strategies are widely adopted in developing new computational methods: 1) making calculations as precisely as possible, 2) making calculations as quickly or simply as possible within a certain range of accuracy. Although the former method is capable to obtain very accurate results, usually it is limited by the size or dimension of the system. The advantage of the latter method is the capability to deal with larger or more realistic systems. In this paper, by persisting the essence of many-body physics, we try to develop a new method to quickly calculate the ground state properties of a quantum many-body system. 

In our previous work\cite{Pan}, the current authors have shown that for one type of matrices, the ground state eigenvector could be approximately determined by the matrix elements. Specifically speaking, the components of the ground state eigenvector are linearly correlated with the sums of matrix elements in corresponding rows. More importantly, this linear relationship holds for larger dimensional matrices with higher accuracy. Yet as a first attempt, we have found that, this linear relationship is not always valid, but only valid for a limited number of special types of matrices, in which the magnitude of diagonal elements are of the same order of off-diagonal ones.\cite{Pan} However, most of many-body Hamiltonian matrices do not have this characteristic. For real quantum many-body matrices, the diagonal elements are often much larger than off-diagonal ones in magnitude. For this situation, the linear relationship is not valid any more.

Although our previous attempt is not valid for all matrices, it does give us two clear hints. First, it is possible to directly establish the relationship between eigenvectors and matrix elements for more general matrices; Second, if this method can be extended to many-body Hamiltonian matrices of practical interests, it will undoubtedly be a very powerful and efficient method to solve high-dimensional matrices. The goal of this paper is to look for a relationship between the ground-state eigenvector and matrix elements for matrices of practical interests. In this paper, we will focus on one kind of matrices, in which all off-diagonal elements are non-positive, but the diagonal elements can take any values.

\section{Technical Details}
We denote a real symmetric matrix by $H$, and its dimension by $N$. Any off-diagonal element $H_{ij}$, corresponding to $i$-th row and $j$-th column, is either negative or zero. The ground-state eigenvector ($|G\rangle$) can be written as $|G\rangle=\sum_ig_i|e_i\rangle$, which is expanded in the orthogonal complete basis ($|e_i\rangle$) with components $g_i$. The sum of elements in $i$-th row ($S_i$) is calculated directly by $S_i=\sum_jH_{ij}$. Both $g_i$ and $S_i$ are re-scaled according to the normalization condition. It is apparent to demonstrate that, any simultaneous change on all diagonal elements with the same amount has no effects on eigenvectors.\cite{Pan} Thus, our investigation will focus on how the distribution width of matrix elements affect the ground-state eigenvector. Furthermore, the effect of matrix density ($\rho$) (e.g., the number of non-zero elements divided by the total number of elements) is also investigated.

Because few methods or theories can be used for current purposes, as in the previous work\cite{Pan}, we follow the spirit in the framework of Big Data analysis and Machine Learning\cite{Nielsen2015,Hinton2006,Ghiringhelli2015}, in which predictions or conclusions are obtained based on vast data sets even though the microscopic mechanism in the certain problem is not yet clear. This strategy has been successfully implemented in the recent studies of many-body quantum systems\cite{Hinton2006,Ghiringhelli2015,Nielsen2015,Carleo2017,Chng2017PRX,Torlai2018,WangLei2016,Cai2018PhysRevB}. In this paper, we have attempted to realize our goal through the systematic analyses of a large number of random matrices, which not only may cover all matrices in principle, but also has widespread applications in physical study\cite{Wigner1955,Bohigas1984PRL,Aaronson2013,Russell2017,Callaway1991PRB,Janssen2000PRE,Zumbuhl2002PRL,Bahcall1996PRL,AndersonSPGS1,AndersonSPGS2,Guhr1998,Deutsch2018Review,PhysRevE.50.888,RevModPhys.53.385,Gomez2011Review,Shen2008PRC}.

For each matrix, the diagonal (off-diagonal) element is produced randomly in the range of [$-\Lambda\times X$, 0] ([$-X$, 0]) respectively, where both $X$ and $\Lambda$ are positive real numbers. Obviously, the value of $\Lambda$ reflects the relative strength between diagonal and off-diagonal elements. In order to make our conclusions as general as possible, we randomly generate a few thousands of matrices for each ($N$, $\Lambda$, $X$) combination. Then, each matrix is directly diagonalized to obtain $g_i$ and $S_i$ for further investigations.

\section{Results and Discussions}

Before carrying out the quantitative performances, it is helpful to have a qualitative picture about how $g_i$ possibly changing with $S_i$. We start with discussions on the two extreme situations. First, for a diagonal matrix, $i.e.$, $\Lambda$ goes to infinite, $g_i$ should be a step function of $S_i$. Only these $g_i$, which corresponds to the minimum $S_i$, have finite values, while all others vanish. Second, for a non-positive random matrix with comparable diagonal elements, $i.e.$, $\Lambda$ is in the order of one, the correlation between $g_i$ and $S_i$ is almost a perfect linear function\cite{Pan}. For finite $\Lambda$, the relationship should be some kinds of function between the linear and step-like functions. Following the above discussions, we speculate that a power function may be one of the most potential candidates. In fact, as we will show, it is indeed the simplest but useful one.

\subsection{random matrices}
\begin{figure}
\includegraphics[width=0.8\textwidth]{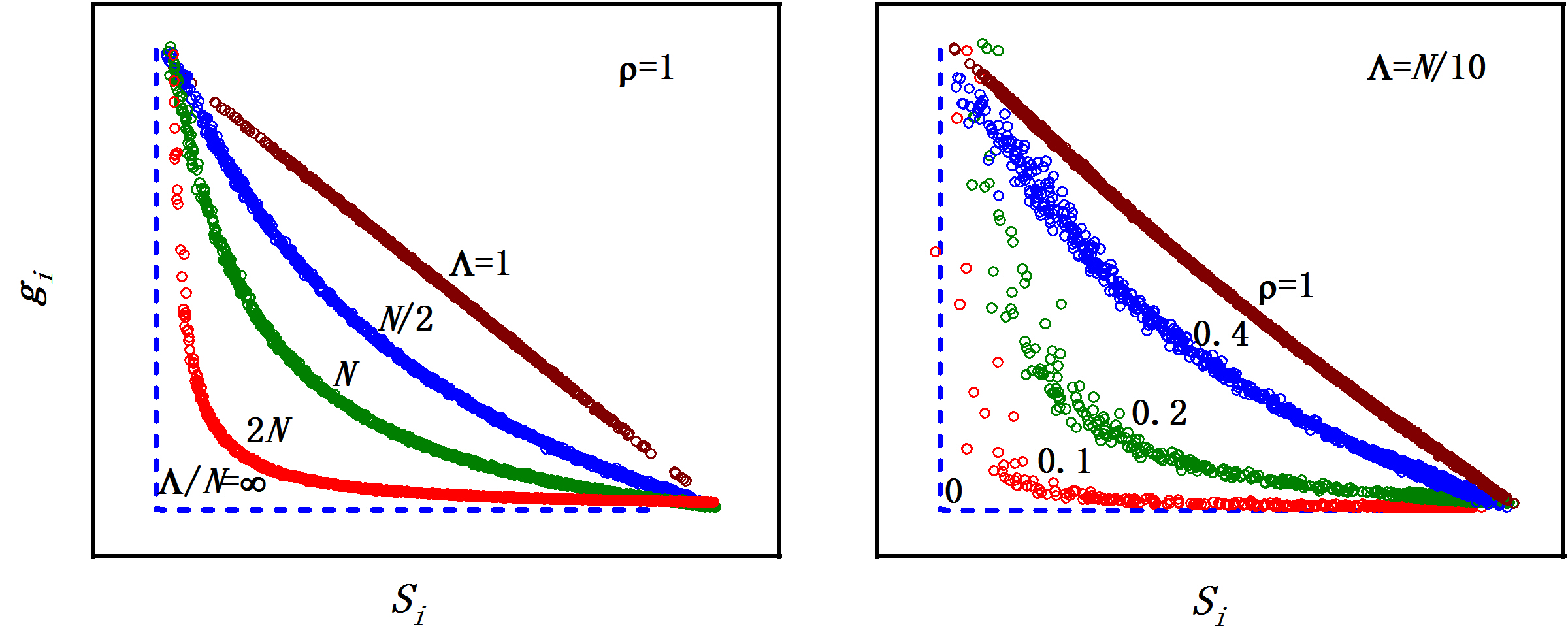}
\caption{\label{D_jieti} (Color online) Elements of the ground-state eigenvector ($g_i$) versus the sum of matrix elements in corresponding row ($S_i$). Left panel: Arbitrary matrix dimension $N$ with different values of $\Lambda/N$ for fixed $\rho=1$. Right panel: Arbitrary matrix dimension for different values of $\rho$ for fixed $\Lambda=N/10$.}
\end{figure}

Fig.\ref{D_jieti} depicts $g_i$ as a function of $S_i$ for different ($N$, $\Lambda$, $\rho$) combinations. From this figure, one can find at least two remarkable features. First, $g_i$ monotonically increases as the decreasing of $S_i$. Although the monotonicity between $g_i$ and $S_i$ holds approximately instead of exactly, the general feature are consistent for all matrices. Second, the correlation presents clear trends with the variation of $\Lambda/N$ and $\rho$. As $\Lambda/N$ goes to infinite or $\rho$ approaches to zero, the matrix is closer and closer to a diagonal matrix, accordingly $g_i$ is more like a step function of $S_i$. By contrary, for $\Lambda/(N\rho)\ll1$, the correlation between $g_i$ and $S_i$ is almost a perfect linear function. 

The result shown in Fig.\ref{D_jieti} provides the possibility to build a specific function between $g_i$ and $S_i$. As mentioned above, the purpose of this paper is, without losing the essential part of physics, to establish a simple relationship between matrix elements and wave functions. As we expected, one of the most potential candidates is a power function, $i.e.$, $g_i=(-S_i)^p+c$, which naturally transits to a linear one as $p=1$ and tends to a step function for large positive $p$. Here both $p$ and $c$ are real numbers. According to the data shown in Fig.\ref{D_jieti}, the values of $p$ and $c$ should be simply determined by the relative values of $\Lambda/N$ and $\rho$. 

\begin{figure}
\includegraphics[width=0.7\textwidth]{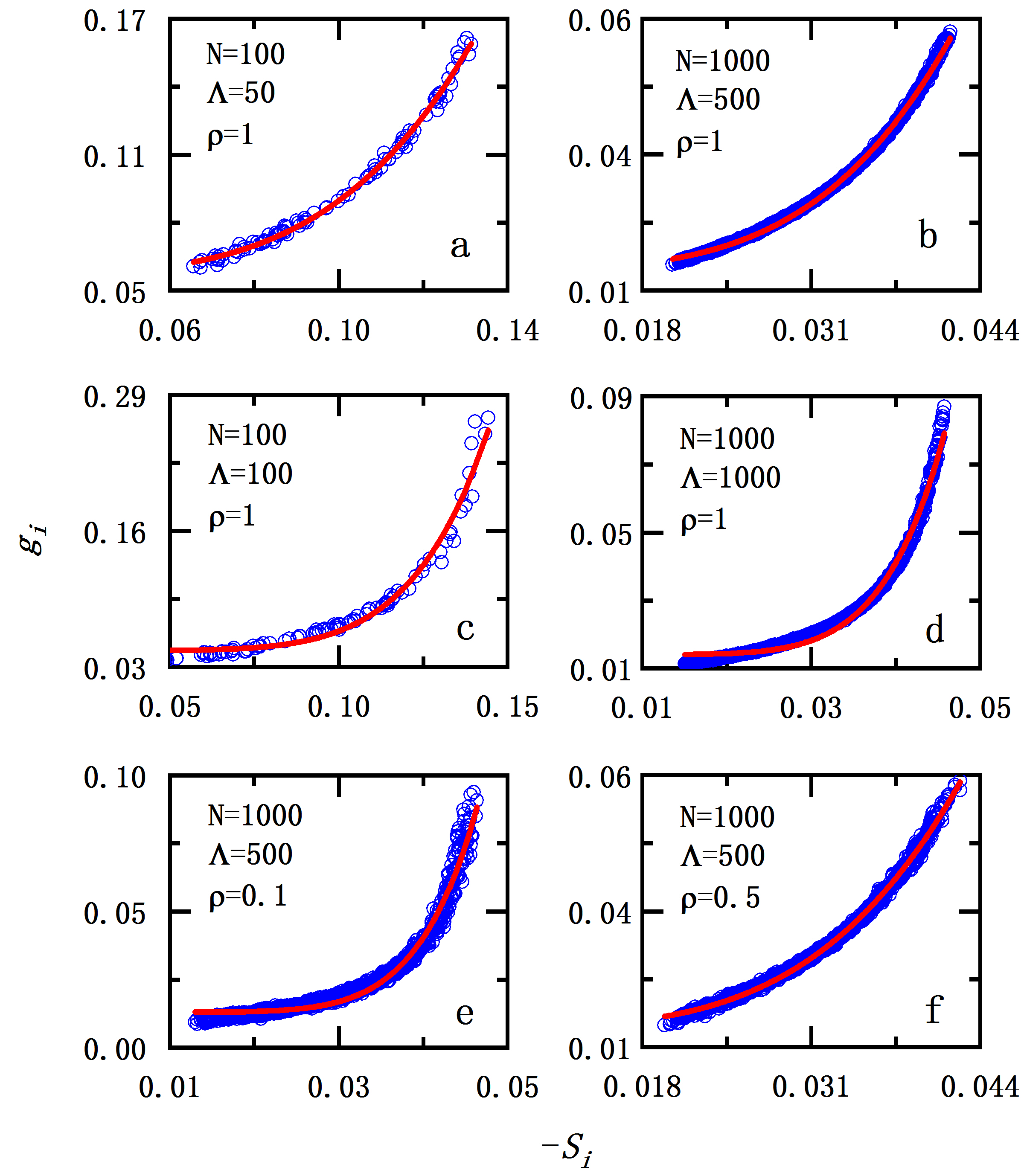}
\caption{\label{D_fit} (Color online) $g_i$ versus $-S_i$ for both dense matrices (a $\sim$ d) with various values of $\Lambda/N$ and sparse matrices (e, f) with various densities, respectively. The fitting curves (solid line) have the form $g_i=(-S_i)^p+c$ with variational parameters $p$ and $c$.}
\end{figure}

We do find that the relation between $g_i$ and $S_i$ can be well fitted by a power-law function $g_i=(-S_i)^p+c$, with $p$ and $c$ varying with both $\Lambda/N$ and $\rho$. All results for both dense and sparse matrices confirm the validity of this power-law function, as shown in Fig.\ref{D_fit}. We can conclude that the power-law function holds for general non-positive  matrices. More importantly, the power-law function just depends on a single parameter, $i.e.$ $\Lambda/(N\rho)$(see below for more details). The simple relationship between $g_i$ and $S_i$ implies a possible new computational method. As will be demonstrated below, this simple form of function indeed captures the physical essence of problems. Although the fitting shown in Fig.\ref{D_fit} can be improved by adding more power terms, we would like to keep a concise form in the following discussion, namely a single power-law function.

\subsection{quantum many-body matrices}
Although there is no rigorous proofs leading to the above power-law relationship by now, since random matrices may cover any specific matrix in principle, we expect that the relationship is hold for  all non-negative Hermitian matrices. To further verify this conjecture, two important models, $i.e.$, the one-dimensional (1D) Hubbard model\cite{Hubbard1963,Zhang1988,Orenstein2000} and the transverse field Ising model\cite{Gennes1963} (or quantum Ising model, equivalently), have been tested. The Hubbard model plays an essential role\cite{Stinchcombe1973,Sachdev_book} in the field of strongly correlated systems, which is one of the most difficult problems in condensed matter physics. Likewise, the quantum Ising model is paradigmatic in our understanding of quantum phase transitions\cite{Sondhi1997,Vojta2003,Sachdev_book}.

The 1D half-filling fermionic Hubbard model contains 4- and 10-sites lattices with anti-periodic- and periodic- boundary conditions respectively. The dimension of Hamiltonian matrices is therefore $36\times36$ and $63504\times63504$, respectively. Its off-diagonal elements are non-positive in the subspace corresponding to $N_{\uparrow}=N_{\downarrow}$. Here $N_{\uparrow}$ ($N_{\downarrow}$) is the total number of spin-up (spin-down) electrons. The on-site coupling strength is chosen to be in the range of $U/t\in[0,8]$. 

For the transverse field Ising model, both 1D and two-dimensional (2D) systems have been considered, the Hamiltonian reads $H_I=-\Gamma\sum_i\hat{\sigma}_i^x-\sum_{<i,j>}\hat{\sigma}_i^z\hat{\sigma}_j^z$, where $\hat{\sigma}_i^x$ and $\hat{\sigma}_i^z$ are Pauli matrices, respectively, and $\Gamma>0$ is a dimensionless parameter. In the basis where $\hat{\sigma}_i^z$ is diagonal, the off-diagonal elements of the Hamiltonian matrix of $H_I$ are constituted by $-\Gamma$ and 0 if the periodic boundary condition is used. For 1D, we have investigated various system sizes with a fixed transverse field strength $\Gamma=2$; for 2D, square lattices of size $4\times4$ with various values of $\Gamma$ are studied. 

\begin{figure}
\includegraphics[width=0.7\textwidth]{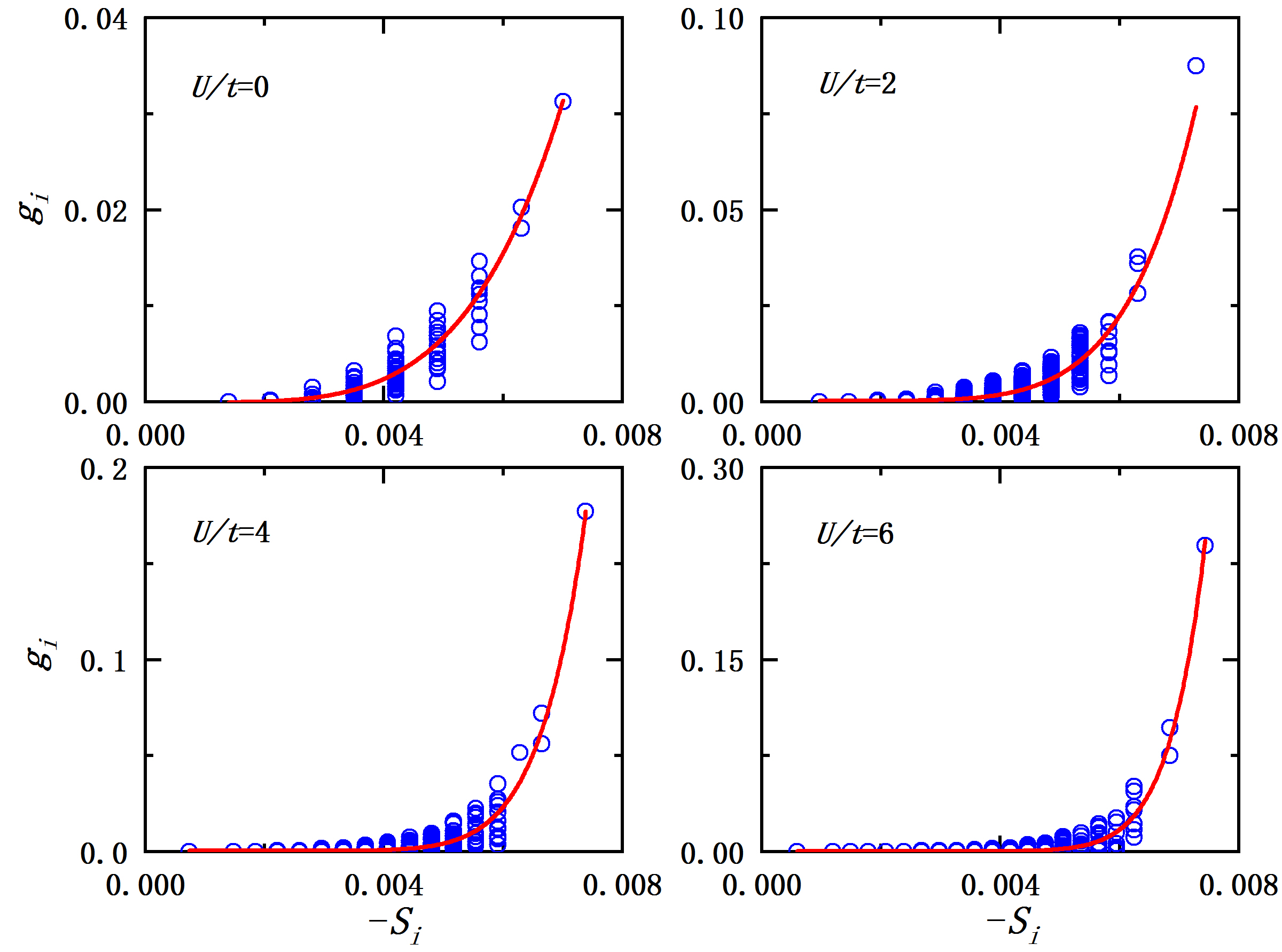}
\caption{\label{Hubbard_scaling} (Color online) $g_i$ versus $-S_i$ for a 1D 10-site Hubbard model with various strengths of the on-site coupling $U/t$. The fitting curves (solid line) have the form $g_i=(-S_i)^p+c$ with variational parameters $p$ and $c$.}
\end{figure}

\begin{figure}
\includegraphics[width=0.7\textwidth]{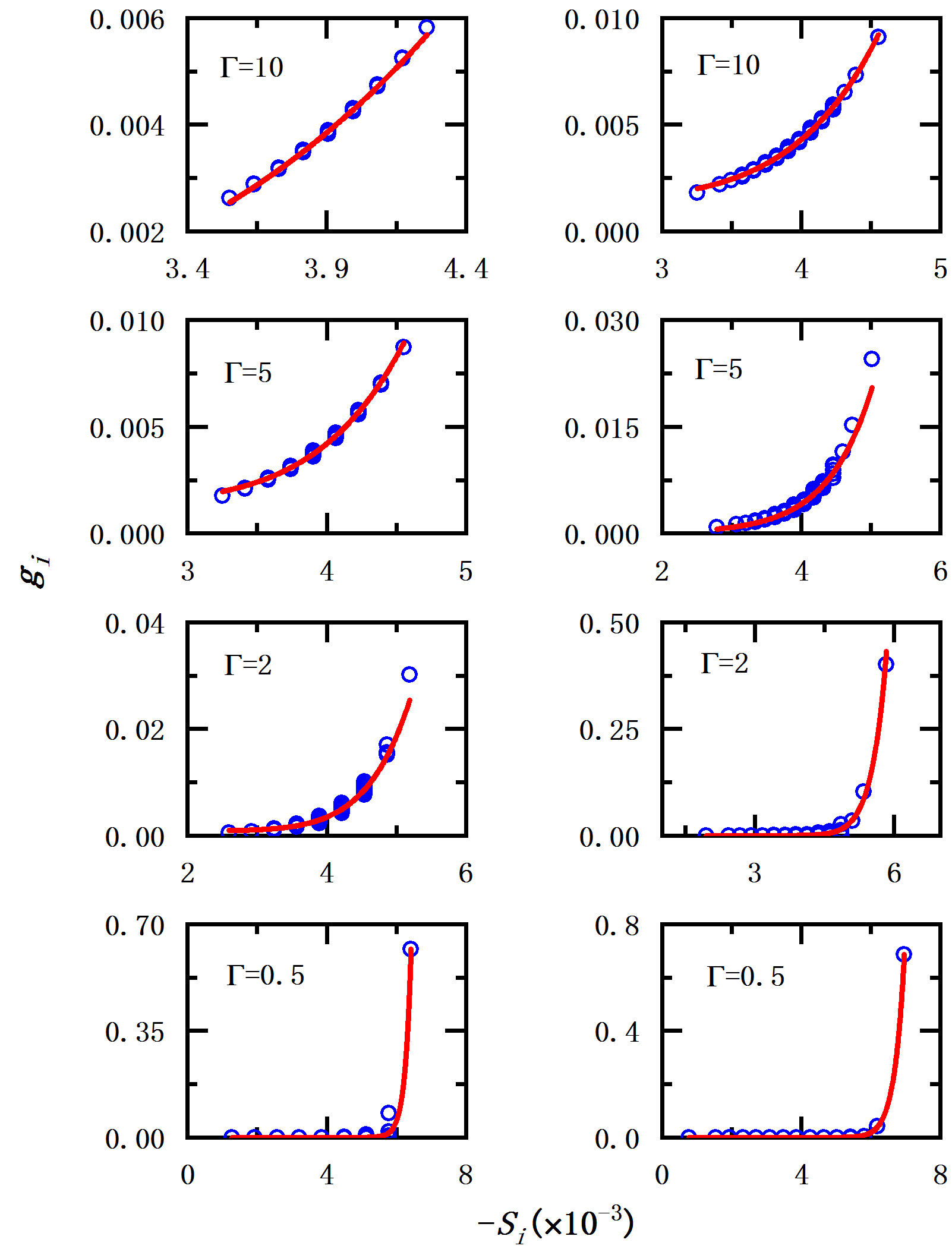}
\caption{\label{Ising_Scaling} (Color online) $g_i$ versus $-S_i$ for transverse field Ising model with various strengths of the transverse field $\Gamma$ for both 16-site chain (left panels) and $4\times4$ square lattice (right panels).}
\end{figure}

Fig. \ref{Hubbard_scaling} presents $g_i$ as a function of $S_i$ for the 10-site Hubbard model. One can see that, as we expected, $g_i$ is an approximate power-law function of $S_i$. Although the $g_i$-$S_i$ curve forms a narrow band, the power-law-like relationship is still maintained. Consistent with the general results for random matrices, the $g_i$-$S_i$ curve becomes steeper for stronger coupling strength, which corresponds to a more diagonal dominated matrix. Similar results are also found in the transverse field Ising model. Fig.\ref{Ising_Scaling} depicts $g_i$ versus $S_i$ for both 1D (left panels) and 2D (right panels) lattices. As expected, $g_i$ decreases with the increase of $S_i$. With the decrease of $\Gamma$, corresponding to diagonal elements being more dominated, the correlation between $g_i$ and $S_i$ deviates from the linear function and tends to step functions. 

The goal of current work is not only to establish the direct relationship between eigenvectors and matrix elements, but also to develop an efficient computational method by using the relationship ($g_i=(-S_i)^p+c$). To do so, the optimized $p$ and $c$ are determined by minimizing the ground state energy ($E$) based on the equation: 
\begin{equation}\label{energy}
E=\frac{\sum_{i,j}g_iH_{ij}g_j}{\sum_ig_i^2}.
\end{equation}
With the optimized $p$ and $c$, the energy, as well as other physical properties of the ground state can be calculated accordingly.

\begin{figure}
\includegraphics[width=0.8\textwidth]{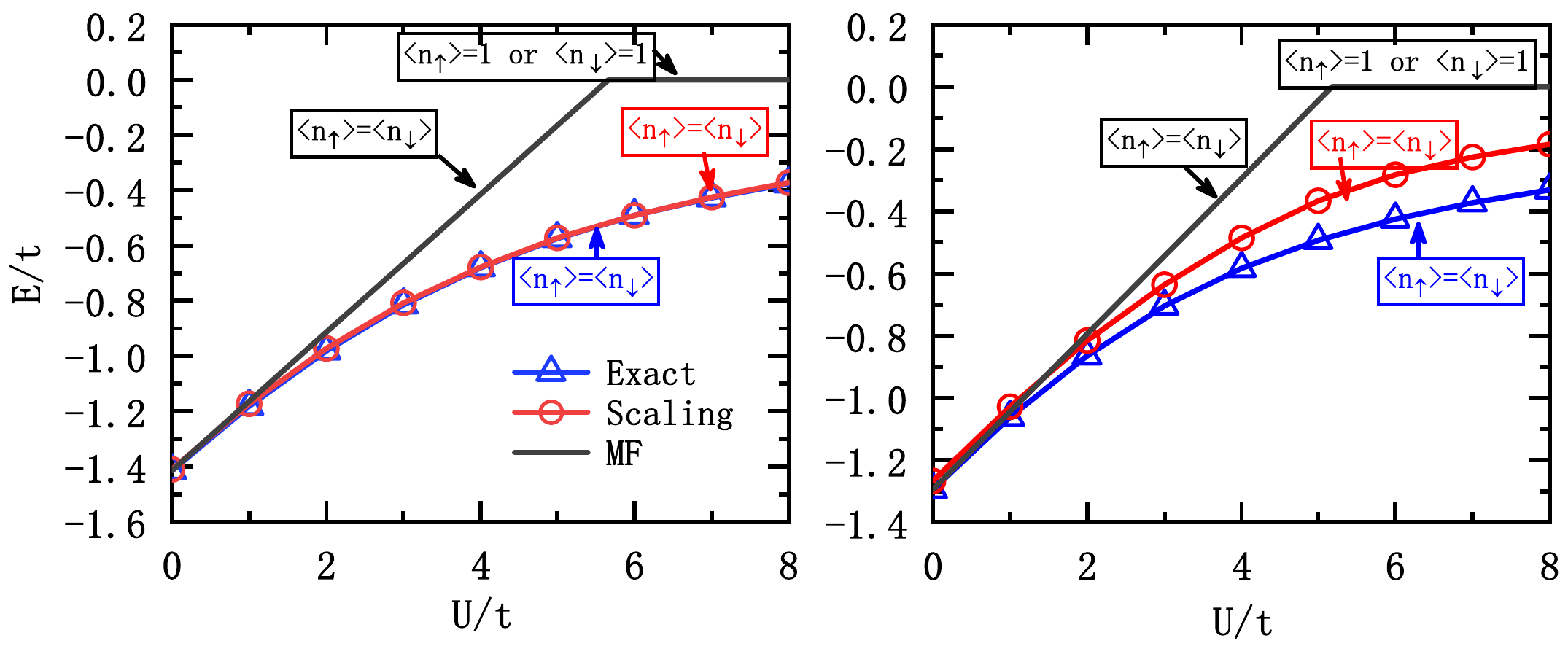}
\caption{\label{Hubbard_Energy2} (Color online) Ground-state energies of the 1D Hubbard model calculated by the power-law relationship (red $\circ$), exact diagonalization method (blue $\triangle$) and mean field theory (black solid line) respectively. Left and right panel show the result for 4-site and 10-site chains, respectively. The strength of coupling varies form $U/t=0$ to $U/t=8$.}
\end{figure}

Fig.\ref{Hubbard_Energy2} presents the energy and magnetization of ground state of 4-site  (left panel) and 10-site (right panel) Hubbard model, in which red circles, blue triangles, and black squares refer the results obtained by current method, the exact diagonalization method, and the mean field theory, respectively. It is well known that, the mean field method performs well in the region of weak coupling, but fails in systems with strong correlation. Especially, there exists a wrong paramagnetic phase for larger $U/t$ in the framework of mean field approximations. In contrast, the current method always predicts the correct magnetic phase for both 4-site and 10-site systems. For the 4-site one, our results are almost identical with exact values, and the accuracy is almost irrelevant with the value of $U/t$. For 10-site one, the difference between exact values and our results increases at larger $U$, but it is still much better than the mean field value. In particular, the magnetic moment well follows the exact values. The errors for 10-site system mainly come from the dispersion of $g_i$ for the same value of $S_i$. We can say that, without losing the essential physics, our current method is indeed a simple and efficient computional method. 

We have also calculated the ground-state energy of the transverse field Ising model by using Eq.(\ref{energy}) with variation on parameter $p$ and $c$. The results are presented in Table.\ref{table1}. We can see that the energy calculated by the scaling relationship is quite accurate. More encouragingly, the relative error $\Delta E/E_{\rm{exact}}$, which is defined by $\Delta E/E_{\rm{exact}}=\frac{E_{\rm{exact}}-E_{\rm{scaling}}}{E_{\rm{exact}}}$, decreases monotonically as the system size increases. 

\begin{figure}
\includegraphics[width=0.8\textwidth]{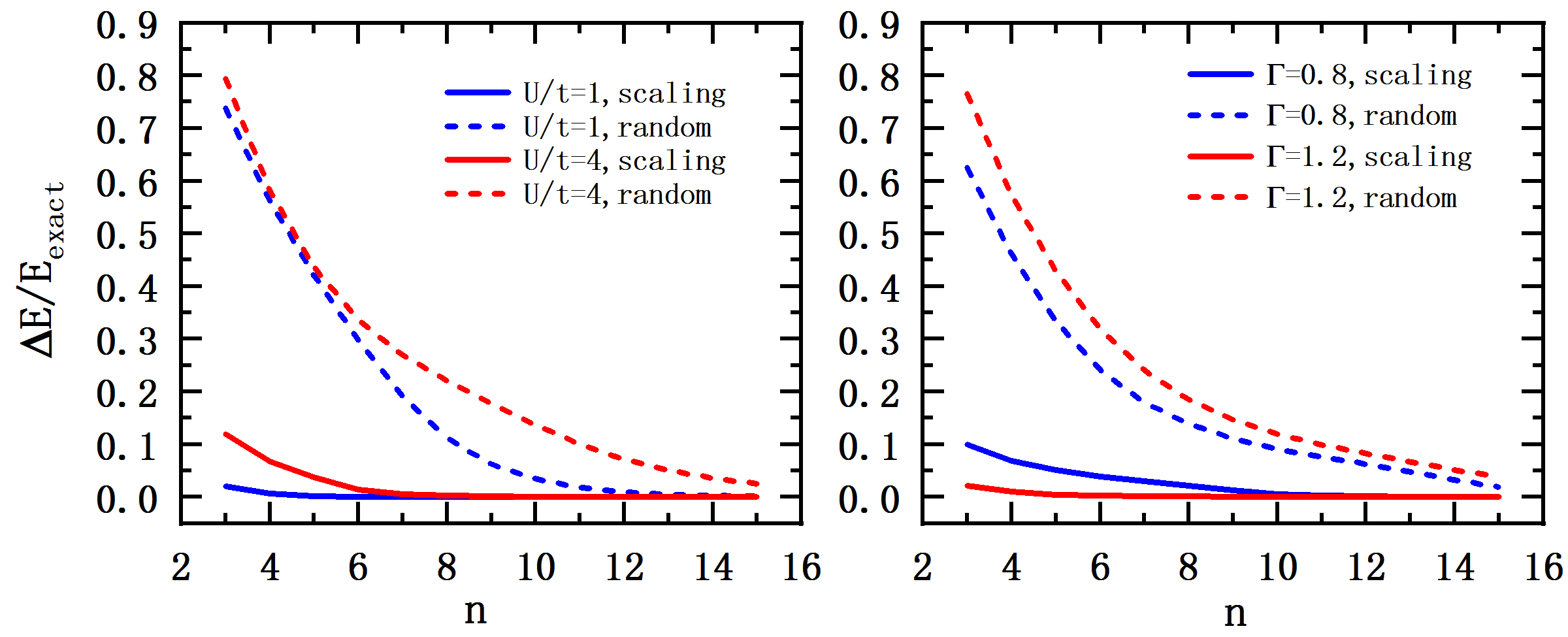}
\caption{\label{Lanczos}The relative error in the ground state energy $\Delta E/E_{\rm{exact}}$ versus the number of iterations $n$ in the Lanczos method. Left panel: results for the Hubbard model on a 10-site chain with $U/t=1$ and 4. Right panel: results for the 16-site transverse field Ising chain with $\Gamma=0.8$ and 1.2. The blue and red lines refer the results in which the initial state is chosen as a random vector and obtained based on the power-law relationship, respectively.}
\end{figure}

Our method can not only efficiently calculate the approximate properties of the ground state, but also provide the optimal initial state for other more accurate methods. Here we will show how the power-law relationship can be used to accelerate the Lanczos method\cite{Lanczos1950}. In the performance of a Lanczos algorithm, the initial state is usually adopted as a random vector. However, we find that the convergence of the Lanczos iteration will be much quicker if the initial state is given according to the power-law relationship of current work. Fig.\ref{Lanczos} shows the convergence of the Lanczos method for both 10-site Hubbard model (left panel) and the 16-site transverse field Ising chain (right panel). Here the 10-site Hubbard model takes $U/t=1$ and 4, and the 16-site transverse field Ising chain takes $\Gamma=0.8, 1$ and $1.2$. In Fig.\ref{Lanczos}, the blue and red lines refer the results in which the initial state is chosen as a random vector and obtained based on the power-law relationship, respectively. The convergence starting from a random state takes at least 15 or more iterative steps, while it costs only a few steps for initial states generated by the power-law relationship. This result not only indicates a practically excellent choice of initial states, but also confirms the fact that the power-law relationship describes ground-state eigenvectors well.

\begin{figure}
\includegraphics[width=0.7\textwidth]{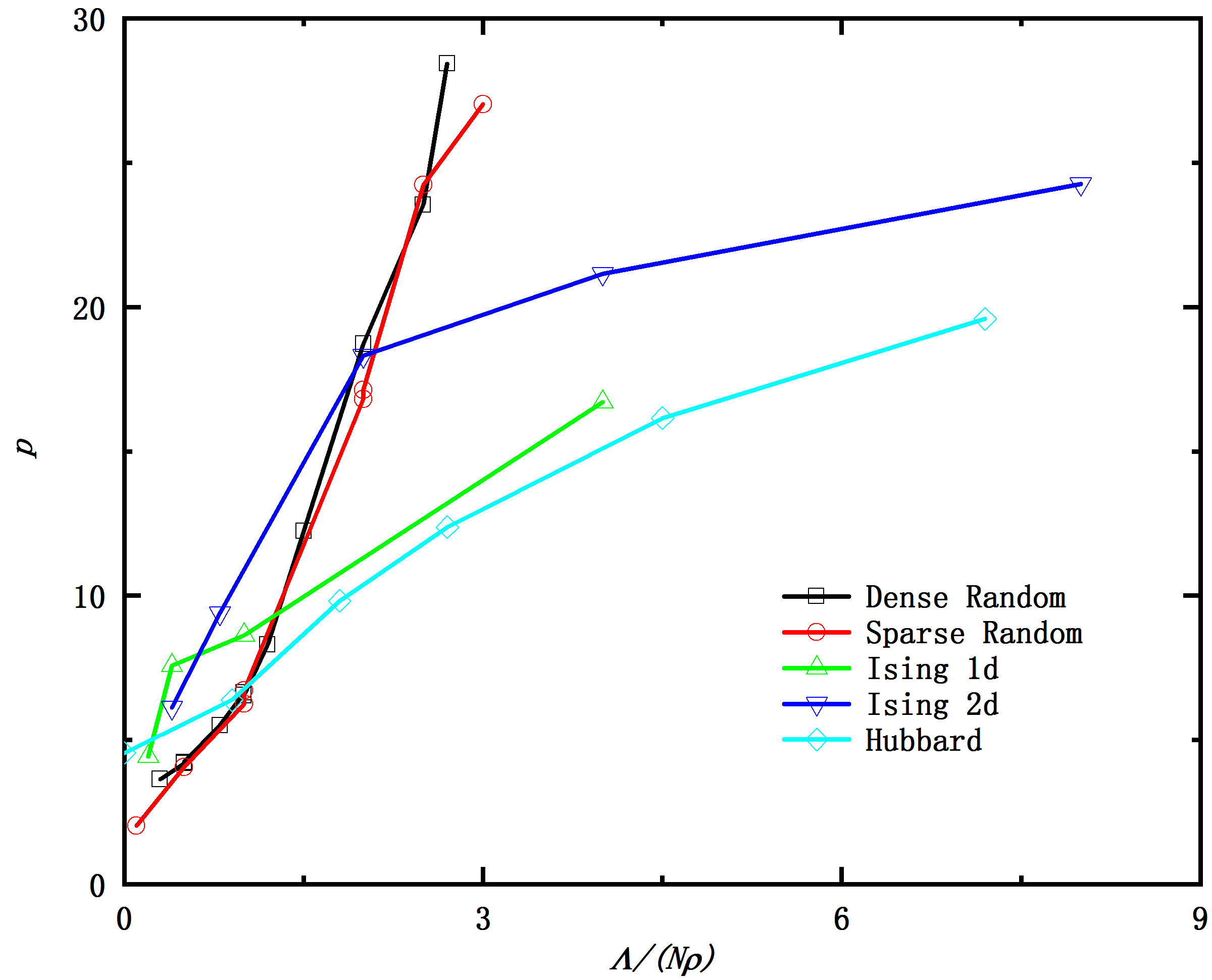}
\caption{\label{exponent}The variation of $p$ as functions of $\Lambda/(N\rho)$ for random matrices with both dense (black $\square$) and sparse cases (red $\circ$) as well as matrices corresponding to Hubbard model (cyan $\diamond$) and transverse Ising model (green $\triangle$ for 1D and blue $\triangledown$ for 2D respectively).}
\end{figure}

Finally, we would like to discuss how the parameter $p$ changes with different matrices. We find that $p$ is a monotone function of $\Lambda/(N\rho)$ for all matrices studied in current work. Fig. \ref{exponent} plots $p$ as functions of $\Lambda/(N\rho)$. One can see that the most important feature is the monotonic increase of $p$ as the increase of $\Lambda/(N\rho)$. In addition to the monotonicity, $p$ is approximately linear except for a possible knee point occurring at a certain value of $\Lambda/(N\rho)$, at which the matrix probably transforms to a diagonal dominated one. The second feature is that the variation of $p$ shows little difference between dense and sparse random matrices. The data shown in Fig. \ref{exponent} confirms our conjecture that $p$ is larger for bigger $\Lambda/(N\rho)$.

\section{Conclusion}
In summary, for any real symmetric matrices with non-positive off-diagonal elements, we have found a simple relationship between the ground-state eigenvector and the matrix elements. The relationship holds well for both random matrices and particular quantum many-body models. According to the simple relationship, we proposed a feasible method to calculate the eigenvector without needing any form of diagonalization.

\section*{Acknowledgements} Project supported by the National Natural Science Foundation of China (Grant No. 11874148). The computations were supported by ECNU Public Platform for Innovation.

\section*{Author contributions statement}
\textbf{Pan Wei:} Conceptualization, Methodology, Software, Writing - Original Draft, Writing - Review and Editing. \textbf{Wang Jing:} Validation, Data Curation, Writing - Review and Editing. \textbf{Sun Deyan:} Methodology, Writing - Review and Editing, Resources, Supervision.

\section*{Declaration of interests}
The authors declare that they have no known competing financial interests or personal relationships that could have appeared to influence the work reported in this paper.

\bibliography{ref}

\begin{table}[H]
\caption{\label{table1}The fitting parameter $p$ and the ground-state energy determined by current method ($E_{\rm{scaling}}$) and exact value ($E_{\rm{exact}}$) for the 1D quantum Ising model on various lattice sizes with $g=2$ and the 2D quantum Ising model on lattice size of $L=4\times4$ with various values of $g$. The relative errors $\Delta E/E_{\rm{exact}}$ are also given for each case.}
\centering
\begin{tabular}{p{2cm}p{2cm}p{2cm}p{2cm}p{2cm}p{2cm}p{2cm}}
%\begin{tabular}{l*{7}{c}}
\toprule[1pt]
                    & \multicolumn{3}{c}{1D}  & \multicolumn{3}{c}{2D}\\
                    \cmidrule(r){2-4} \cmidrule(r){5-7}
                    & $L=8$       & $L=12$     &  $L=16$     & $g=5$      & $g=1$      &  $g=0.5$\\
\hline
$p$                 & 2.1378      & 3.2870     &  4.4359     &  6.0199    & 21.4033    &  24.4341\\
$E_{\rm{scaling}}$  & -2.121003   & -2.122308  & -2.122975   & -5.1030    & -2.113265  & -2.029381\\
$E_{\rm{exact}}$    & -2.126907   & -2.127083  & -2.127089   & -5.106174  & -2.125662  & -2.031291\\
$\Delta E/E_{\rm{exact}}$    & 0.00278  & 0.00224  & 0.00193 & 0.000622   & 0.00583    & 0.000940\\
\bottomrule[1pt]
\end{tabular}
\end{table}
\clearpage

\end{document}